\documentclass[reprint,superscriptaddress,onecolumn]{revtex4-2}
\pdfoutput=1 

\usepackage{natbib}
\usepackage{lmodern}
\usepackage{graphicx}
\usepackage{mathtools}
\usepackage{mathrsfs, amsmath}
\usepackage{amsfonts}
\usepackage{amssymb}
\usepackage{lipsum}
\usepackage[export]{adjustbox}
\usepackage{cancel}
\usepackage{dcolumn}
\usepackage{bm}
\usepackage{hyperref}
\hypersetup{linktocpage,colorlinks,citecolor={blue},pdfdisplaydoctitle=true,pdfpagemode=UseOutlines,bookmarksnumbered=true}
\usepackage{mathrsfs,dsfont}
\usepackage{color}
\usepackage{wrapfig}
\usepackage{comment}
\usepackage{lipsum}  
\usepackage[normalem]{ulem}

\begin{document}
\onecolumngrid

\preprint{APS/123-QED}

\title{Transition from viscous fingers to foam during drainage in heterogeneous porous media}

\author{Federico Lanza}\email{federico.lanza@universite-paris-saclay.fr}
\affiliation{
Universit\'e  Paris-Saclay,  CNRS,  LPTMS,  91405,  Orsay, France
}%
\affiliation{
PoreLab, Department of Physics,\\ Norwegian University of
Science and Technology, N-7491, Trondheim, Norway
}%

\author{Santanu Sinha}\email{santanu.sinha@ntnu.no}
\affiliation{
PoreLab, Department of Physics,\\ Norwegian University of
Science and Technology, N-7491, Trondheim, Norway
}%

\author{Alex Hansen}\email{alex.hansen@ntnu.no}
\affiliation{
PoreLab, Department of Physics,\\ Norwegian University of
Science and Technology, N-7491, Trondheim, Norway
}%

\author{Alberto Rosso}\email{alberto.rosso@universite-paris-saclay.fr}
\affiliation{
Universit\'e  Paris-Saclay,  CNRS,  LPTMS,  91405,  Orsay, France
}%

\author{Laurent Talon}\email{laurent.talon@universite-paris-saclay.fr}
\affiliation{
Universit\'e  Paris-Saclay,  CNRS,  FAST,  91405,  Orsay, France
}%

\date{\today}

\begin{abstract}
We investigate the behavior of drainage displacements in heterogeneous porous media finding a transition from viscous fingering to foam-like region. A pore network model incorporating the formation of blobs is adopted to study this phenomenon. By imposing a pressure difference between the inlet and outlet, we observe that the displacement pattern undergoes a significant transition from a continuous front of growing viscous fingers to the emergence of foam, which develops and propagates until breakthrough. This transition occurs at a specific distance from the inlet, which we measure and analyze as a function of the viscosity ratio and the capillary number, demonstrating that it follows a non-trivial power-law decay with both the parameters. Moreover, we discuss the relationship between the evolution of the total flow rate and the local pressure drop, showing that the foam developed reduces global mobility. We observe that foam is formed from the fragmentation of viscous fingers beneath the front, and this instability mechanism is connected with fluctuations of the local flow rate, which we analyze both in the viscous fingering region and in the foam region.
\end{abstract}

\maketitle

\section{\label{sec:introduction}Introduction}

Simultaneous flow of multiple immiscible fluids in porous media, named as multi-phase flow \cite{b88,b17,ffh22}, are involved in a wide range of industrial and geophysical applications. A few important examples for instance are the understanding of the water cycle, the transport of pollutants in soils, oil recovery, geothermal energy extraction and carbon dioxide sequestration. In two-phase flow, when a more wetting fluid displaces a less wetting fluid inside a porous medium, the flow is referred as imbibition, whereas the opposite case, when a less wetting fluid displaces a more wetting fluid, it is called drainage. The study we present here deals with drainage, and we will refer the more- and less-wetting fluids as wetting and non-wetting fluids respectively. Depending on the flow parameters and system properties, the displacement process produces fronts with distinct shapes which characterize the underlying flow mechanism. There can be stable displacement with flat front \cite{bbo15} when a fluid of high viscosity invades a fluid with much lower viscosity under a viscous pressure drop, or, there can be more complex fingering patterns with fractal structures \cite{f88}. Two essential mechanisms control the shape of the fingers. First, there is viscous fingering \cite{mfj85,h87,sss22}, which is an instability that occurs during the fast displacement of a low-viscosity fluid injected into a more viscous fluid \cite{st58}. Structure of such fingers are strongly analogous to diffusion limited aggregation (DLA) \cite{ws81}, both of which obey Laplacian growth. The second is the capillary fingering \cite{lz85} which appears during slow displacement when the dynamics is controlled by the disorder in the capillary forces related to the pore size distribution. Structure of capillary fingers are different than the viscous fingers and can be modeled by invasion percolation theory \cite{ww83,w86}. A detailed study on the crossover between different regimes due to the competition of the viscous and capillary effects can be found in Lenormand et al. \cite{ltz88,lz89}.

Both viscous fingering and percolation theory share the common feature that the invading fluid remains connected. However, in many two-phase flow scenarios, it is not uncommon that the invading fluid breaks and becomes disconnected. The breakage of the phase in porous media has also been widely investigated \cite{payatakes84}. To summarize, three mechanisms are usually invoked. First, there is the Roof snap-off \cite{roof70, Rossen03}. When the tip of the invading fluid exits a constriction, the curvature at the tip decreases while the curvature inside the constriction remains high. This induces a pressure difference that might create a back-flow of the defending fluid which break the invading one. Another important mechanism is the pinch-off, where a continuous fluid filament breaks up into smaller parts due to the Rayleigh-Plateau instability \cite{tomotika35}. The last mechanism, the so-called dynamic breakup \cite{payatakes82}, occurs when the tip of a phase encounter a junction of two pores, where it might split to invade both and disconnect then the other phase. Despite the abundant literature on the subject, it remains difficult to establish the effective role and relevance of these mechanisms in both generating and propagating foam in porous media. For example, in the last years, pore network simulations have shown \cite{Chen2006} that strong foam generation can occur without the necessity of Roof snap-off mechanism.

Phase break-up is of particular interest to generate foam inside the porous structures \cite{Lake14}. A particular application of foam flow is, for instance, the enhanced oil recovery. As discussed above, displacing a more viscous fluid (e.g oil) by a less viscous one (e.g water) creates unstable finger-like preferential paths. In this way, a certain amount of fluid is left in place, altering significantly the recovery.
One method to circumvent this problem is to inject foam which increases significantly the viscosity of the displacing fluid \cite{falls89}. A consolidated technique consists in injecting gas and water with surfactant, to create the foam in situ \cite{ransohoff88}. The main objective here is to generate \textit{strong} foam with low mobility (foam with smaller blob sizes with large number of lamellae), in opposition to the \textit{weak} one with high mobility (large blobs with fewer lamellae) \cite{Gauglitz02, dicksen02,dsk07}.

It can be thus noted that the two types of displacement mentioned, continuous front and foam flow, have been widely studied separately in the literature. One would expect, however, that there might be a transition between these two regimes. Notably, it was early reported experimentally and numerically that a condition for generating foam in situ is to apply a minimum flow or pressure gradient \cite{ransohoff88, Rossen90, dicksen02}. On the other hand, it may be argued that, if the flow is subjected to an imposed pressure, the development of capillary or viscous fingering leads to an evolution of the flow rate, and of the homogeneity of the the pressure gradient as well. It is not unlikely that a fingering regime evolve toward a foam displacement. In a recent paper, Eriksen et al.\ \cite{Eriksen2022} observe experimentally a transition from compact displacement to viscous fingering. They find that, in a radial geometry with imposed pressure drop, the invasion front is initially stable and the invading pattern presents an intense blob dynamic typical of foam, but, after a certain radius, the front adopts a viscous-fingering-like shape. It is perhaps useful to point out that, when injecting the fluid at a constant flow rate in the radial geometry, the velocity of the front tends to decrease naturally due to the conservation of mass. If a critical velocity is necessary for the generation of foam, it is not excluded that the latter stops after a certain distance.\\
In this connection, it should be noted that on the Darcy scale, where the porous medium may be regarded as a continuum, the fractal structures generated by continuous front flow e.g., viscous fingers will have a zero measure.  It will be foam-like structures, which are not fractal, that result in variations of the saturation at such scales.\\ 
The objective of this work is to better understand the transition between the two types of displacements, continuous flow and foam flow. We show that, when injecting a low-viscous fluid into a porous medium filled with a high viscous fluid at constant global pressure drop in a rectangular system, under certain conditions, the system shows a transition from viscous fingering regime to a compact foam flow. In this regime, the invasion starts with viscous fingers. However, as the front advances towards the outlet, the fingers tends to break into small droplets, developing foam after a certain distance from the inlet. We implement a dynamical pore-network model that is capable of modelling both the classical displacement fronts and blob generations. To characterize this transition, we implemented a method for distinguish the region of viscous fingering from the region of foam development and propagation, measuring the distance of the crossover line from the inlet. Moreover, we showed the effect of the foam formation on both global quantities, namely the total flow rate, and local quantities, like the pressure gradient and the local flow rate. Finally, the origin of this transition is discussed.

\section{Pore Network Model}
\begin{figure}
    \centering
    \includegraphics[scale=0.12]{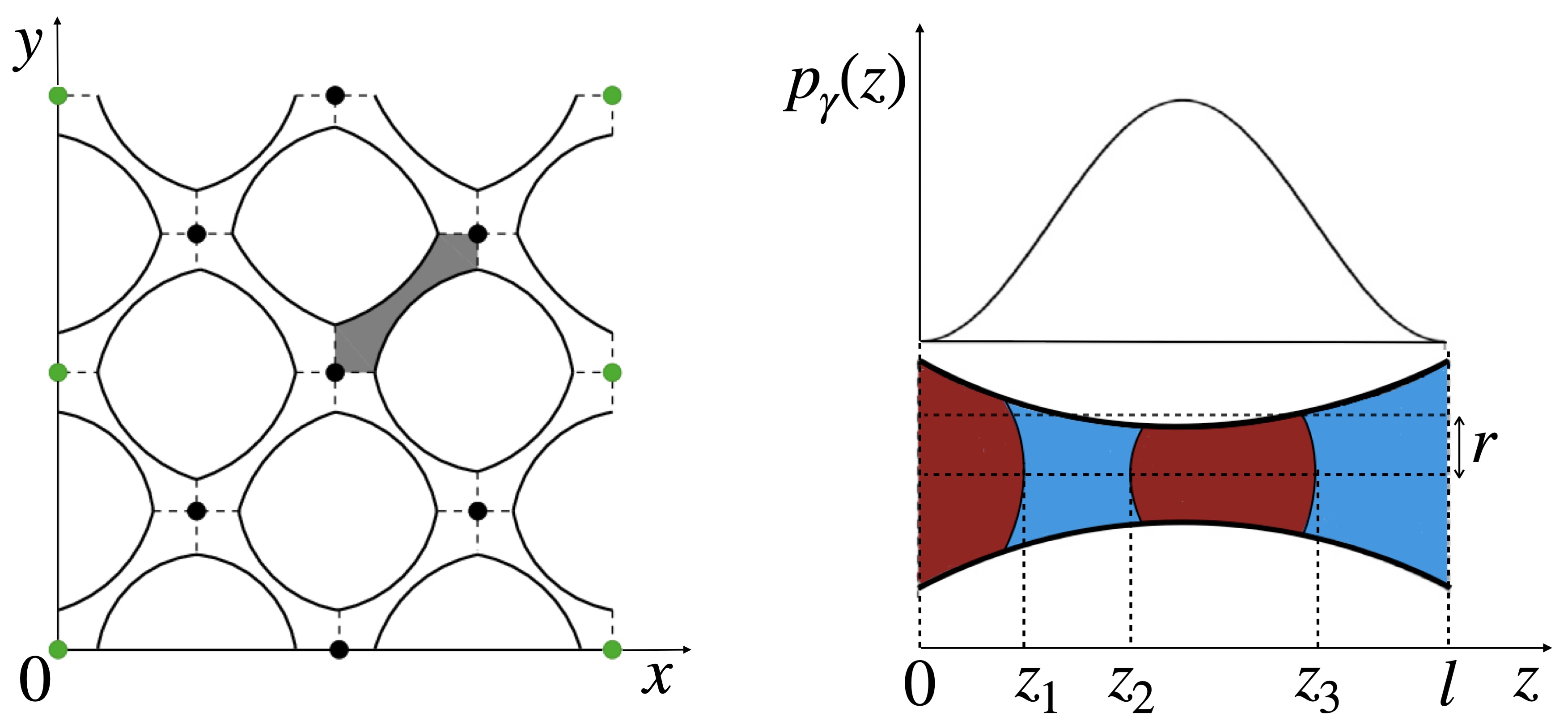}
    \caption{Left: Illustration of a pore network made by $N_x\times N_y = 4\times 4$ links. The hour-glass shaped links are connected to each other at nodes denoted by circular dots. Because of the periodicity, green dots at the same height represent the same node. One of the links is colored gray. Right: Illustration of a single link filled with two fluids, separated by different interfaces (menisci). The blue and red color represent the wetting and non-wetting fluid, respectively. The capillary pressure drop for an interface, expressed by equation \eqref{eq:YoungLaplace}, is shown above the link, where $p_\gamma(z)$ is plotted as a function of the interface position $0 < z < l$.}
    \label{fig:sketch}
\end{figure}

Simulations were carried out in the framework of dynamic pore-network modeling \cite{Blunt2001}. The pore network we consider here has an underlying geometry of a regular square lattice of $N_x\times N_y$ links as shown in Figure \ref{fig:sketch}, which is tilted by an angle of $45^\circ$ with the direction of the global pressure drop. The network consists of {\it composite links}, which means that each link contains a narrow pore throat in between two wider pore bodies. This is modeled by having links with varying radius along its length, similar to an hourglass shape. The total porous space of the network is therefore contained by all the links, and the nodes represent only the positions of the intersections of the links. We consider all links with identical length $l$, and presenting axial symmetry. The disorder is then introduced in the charateristic radius, $r$, of the links. Two different distributions for $r>0$ are considered:
\begin{itemize}
\item Uniform distribution
\begin{equation}
    \Pi(r) =\begin{dcases} 1/a & \text{if}\; r\in \left[ \overline{r} - a/2, \overline{r} + a/2 \right]\\
0 & \text{otherwise}
\end{dcases}\!;
\label{eq:Pi_uniform}
\end{equation}
\item Rayleigh distribution
\begin{equation}
\Pi(r) = \frac{r}{\overline{r}^2}e^{-\frac{r^2}{2\overline{r}^2}};
\label{eq:Pi_Rayleigh}
\end{equation}
\end{itemize}
here, $\Pi(r)$ is the corresponding probability density function, $\overline{r}$ is the average radius and $a$ is the interval width for the uniform distribution.

We restrict to the creeping laminar-flow regime where the Poiseuille law is valid. In that case, if the link is filled with one Newtonian fluid of viscosity $\mu$, the flow rate $q_{ij}$ inside a link in between two nodes $i$ and $j$ depends on the pressure drop, $p_i - p_j$, between the two nodes,
\begin{equation}
\label{eq:PoiseuilleLaw}
   \displaystyle
   q_{ij} = \frac{\pi r_{ij}^4}{8 \mu l} \left(p_i - p_j \right), 
\end{equation}
where we assumed that the radius does not deviate too much from its characteristic value $r_{ij}$ of the link. In case of two immiscible fluids present in the same link, Equation \ref{eq:PoiseuilleLaw} needs to be modified \cite{w21}. First, the viscosity $\mu$ will be the effective viscosity $\mu_{ij} = \mu_{n}S_{ij} + \mu_{w}(1-S_{ij})$. Here $S_{ij}$ is the local non-wetting saturation, namely the fraction of the link length occupied by the non-wetting fluid, while $\mu_{n}$ and $\mu_{w}$ are respectively the non-wetting and wetting viscosities. Second, there is a capillary pressure drop $p_\gamma$ across the meniscus separating the two fluids that must be taken into account. As the links of the network have a varying radius along their length, $p_\gamma$ will depend on the position of a meniscus. For the converging-diverging type of an hourglass-shaped link, we model the variation of the capillary pressure with the position $0<z<l$ of a meniscus by a modified Young-Laplace law \cite{amh98,shb13},
\begin{equation}
\label{eq:YoungLaplace}
   \displaystyle   
   \left| p_\gamma(z)\right| = \frac{2\gamma}{r_{ij}}\left[1-\cos \left(\frac{2\pi z}{l}\right) \right] \;,
\end{equation}
where $\gamma=\hat\gamma\cos\theta$, $\hat\gamma$ being the surface tension between the two fluids and $\theta$ the contact angle between the meniscus and the link wall, which is assumed not to vary during the motion. With these two modifications, Equation \ref{eq:PoiseuilleLaw} for a number of $m$ menisci inside a link can be generalized as,
\begin{equation}
    \label{eq:qVSdp}
    \displaystyle 
    q_{ij} = \frac{\pi r_{ij}^4}{8\mu_{ij}l} \left[p_i - p_j - \sum_{k=1}^m  p_\gamma\left(z_k\right)\right] \;,
\end{equation}
where the summation is over all the interfaces $k=1,\dots,m$, inside the link, taking into account the direction of the capillary forces.

We simulate the drainage displacement where a less-viscous, non-wetting fluid invades a network filled with a more-viscous, wetting fluid. This is done by filling the network completely with the wetting fluid initially, and then injecting the non-wetting fluid at one edge of the system, marked as inlet ($y=0$). The opposite edge of the system is marked as outlet ($y=N_y$), through which fluids leave the network. As we perform the simulations at a constant pressure drop $\Delta P$, we impose a fixed pressure value $P_{in} = \Delta P$ at all of the inlet nodes and  $P_{out} = 0$ at all of the outlet nodes. This creates an overall global pressure drop $\Delta P$ in the direction of the inlet edge to the outlet edge of the network. The two lateral edges of the network parallel to the direction of pressure drop are connected using periodic boundary condition.

As both the fluids are incompressible, the net volumetric flux of the fluids at any given node will be zero for every time $t$ during the invasion process. This is analogous to the first Kirchhoff law for the electrical current, and can be expressed for every node $i$ as
\begin{equation}
    \label{eq:Kirchhoff}
    \displaystyle
    \sum_{j\in\text{ngb}(i)} q_{ij}(t) = 0 \;,
\end{equation}
where ${j\in\text{ngb}(i)}$ are the neighboring nodes connected to the node $i$ by links. This provides a closed set of linear equations which, once solved, allows to compute both the local node pressures $\{p_i(t)\}$ and the flow rates $\{q_{ij}(t)\}$ for the whole network. From the local flow rates we update the positions of each menisci in the system, displacing it by a distance
\begin{equation}
    \label{eq:Deltax}
    \displaystyle
    \Delta z_{ij} = \frac{\Delta t\,q_{ij}}{\pi r_{ij}^2}
\end{equation}
in the direction of the local flow. Here the time step $\Delta t$ is chosen in such a way that the largest displacement of any meniscus in any link does not exceed $0.1\, l$ in one time step.

To distribute fluids from links to their neighboring links at the links intersections or nodes, we consider an algorithm that does not impose any restriction on the blob sizes inside any link, and for which the blob sizes are determined by the dynamics of the flow \cite{Sinha2021}. This makes it possible for the model to generate not only the continuous capillary or viscous fingers but also foams with smaller blobs. The model can therefore capture the transition from fingering to foam formations while changing external flow parameters, and no alteration in the fluid distribution algorithms is necessary. The algorithm first calculates the total volume of fluids $V_i=-\sum_j q_{ij}\Delta t$ that each node $i$ receives from the {\it incoming} neighboring links, namely the links for which $q_{ij}<0$ for a node $i$ according to equation \ref{eq:qVSdp}. The individual values of the wetting and non-wetting volumes $V_i^w$ and $V_i^n$ that the node receives from the incoming links are calculated from the displacements of the menisci following equation \ref{eq:Deltax}. The two fluid volumes are then redistributed to the {\it outgoing} neighboring links, the links for which $q_{ij}>0$ for any node $i$. The redistribution follows an impartial rule where the ratio between the total injected volumes of fluids $V_{ij}$ in different outgoing links is equal to the ratio between the flow-rates $q_{ij}$ in those links, and the ratio between the volumes of wetting ($V_{ij}^w$) and non-wetting ($V_{ij}^n$) fluids in each individual outgoing link is proportional to the incoming wetting and non-wetting volumes $V_i^w$ and $V_i^n$ in the distributing node. This is done by creating new wetting and non-wetting blobs of volume $V_{ij}^w = q_{ij}\Delta tV_i^w/V_i$ and $V_{ij}^n = q_{ij}\Delta t V_i^n/V_i$ respectively in every outgoing link of node $i$. The order of the new wetting and non-wetting droplets are chosen arbitrary. Furthermore, when the number of the blobs exceeds a maximum limit in a link, we merge two nearest blobs keeping the volume conserved, but without detaching any blob attached to a node which may be a part of a cluster spanned over several links. 

In summary, at every time step $\Delta t$, we calculate the local pressures $p_i$ for each node and flow rates $q_{ij}$ in each link by solving equations \ref{eq:qVSdp} and \ref{eq:Kirchhoff}, then update the positions of the menisci using equation \ref{eq:Deltax}. The fluids are then exchanged between different links which in general alters the local saturation $S_{ij}$ in the links, as well as the number and positions of the menisci. This necessitates the linear system of equations \eqref{eq:qVSdp} and \eqref{eq:Kirchhoff} to be solved again in the next time step.

As discussed in the introduction, the main competing mechanisms that control the flow characteristics in two-phase flow are the ratio between the viscosities of the two fluids, called the viscosity ratio, and the ratio between the viscous and the capillary forces at the pore level, know as the capillary number. The viscosity ratio is defined as $M=\mu_n/\mu_w$, whereas the capillary number is generally defined for the flow driven under the constant flow rate $Q$ as \cite{Lenormand1988,fms97,amh98},
\begin{equation}
    \label{eq:CaQ}
    \displaystyle
    \text{Ca}_Q = \frac{\mu_w Q}{\gamma A}\;,
\end{equation}
where $A = N_x \pi \overline{r}^2$ is the average cross-sectional area of the pore network. However, when the system is driven under a constant pressure drop $\Delta P$ as we are studying here, the total flow rate $Q$ varies with time. The capillary number is therefore defined as a function of the pressure drop \cite{lmm05},
\begin{equation}
\text{Ca}_P = \frac{{\Delta P}/{N_y}}{{2\gamma}/{\overline{r}}}.
\label{eq:CaP}
\end{equation}
where $\Delta P/N_y$ is the average pressure drop across one link and $2\gamma/\overline{r}$ is the typical capillary pressure drop for a meniscus in a link. These two dimensionless numbers fully characterize the invasion displacement. This means in particular that different simulations with the same values of $M$ and $\text{Ca}_P$ will produce invasion patterns which are statistically equivalent. In the Supplementary Material we show the validity of this statement.

\section{Transition from viscous fingering to foam}
\begin{figure}
    \centering
    \includegraphics[scale=0.13]{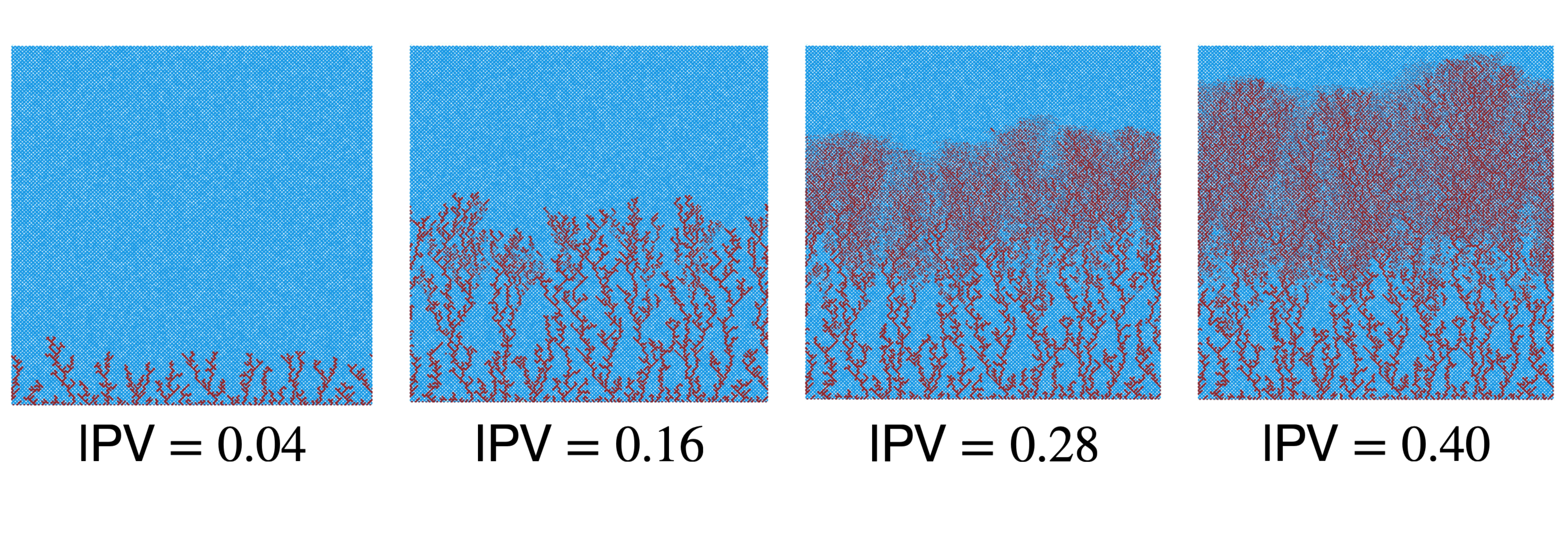}
    \caption{Snapshots at different growing values of the injected pore volume (IPV). The blue and red color represent the wetting and non-wetting fluid, respectively. We observe a transition from viscous fingering to foam at a certain distance from the inlet. For this simulation we set $\text{Ca}_P = 0.25$, $M=10^{-2}$, and $r$ generated according to the uniform distribution \eqref{eq:Pi_uniform} with $\overline{r} = 0.25\, l$ and $a=0.15\, l$.} 
    \label{fig:timeEvolution_snapshot}
\end{figure}
Figure \ref{fig:timeEvolution_snapshot} displays the evolution in time of a typical invasion front with $M=10^{-2}$, $\text{Ca}_P = 0.25$ and $r$ uniformly distributed. Unless indicated
otherwise, the results presented will correspond to a lattice of size $N_x \times N_y = 200 \times 200$. Here, we describe the evolution of our system using the normalized injected pore volume $\text{IPV} = V_{\text{inj}}/V_{\text{tot}}$, where $V_{\text{inj}}$ is the volume of the injected fluid and $V_{\text{tot}}$ is the volume of the total pore space of the network. Initially, the non-wetting phase enters in the porous medium and exhibits the common viscous fingering occurring at high flow rate \cite{ltz88}. At this early stage, the non-wetting fluid remains continuous in the absence of any breakage. Also, we note that the invaded pores are never occupied again by the wetting phase. However, as the fingers advances through the network, we observe that the continuous non-wetting fingers eventually break up and the invading pattern exhibits a very different foam-like structure composed of many small droplets.
Notably,  this transition between the continuous viscous fingers to the foam regime appears at a certain distance $\Lambda$ from the inlet that depends on both the capillary number $\text{Ca}_P$ and viscosity ratio $M$. We will analyse this in detail in the following section.

\begin{figure}
    \centering
    \includegraphics[scale=0.34]{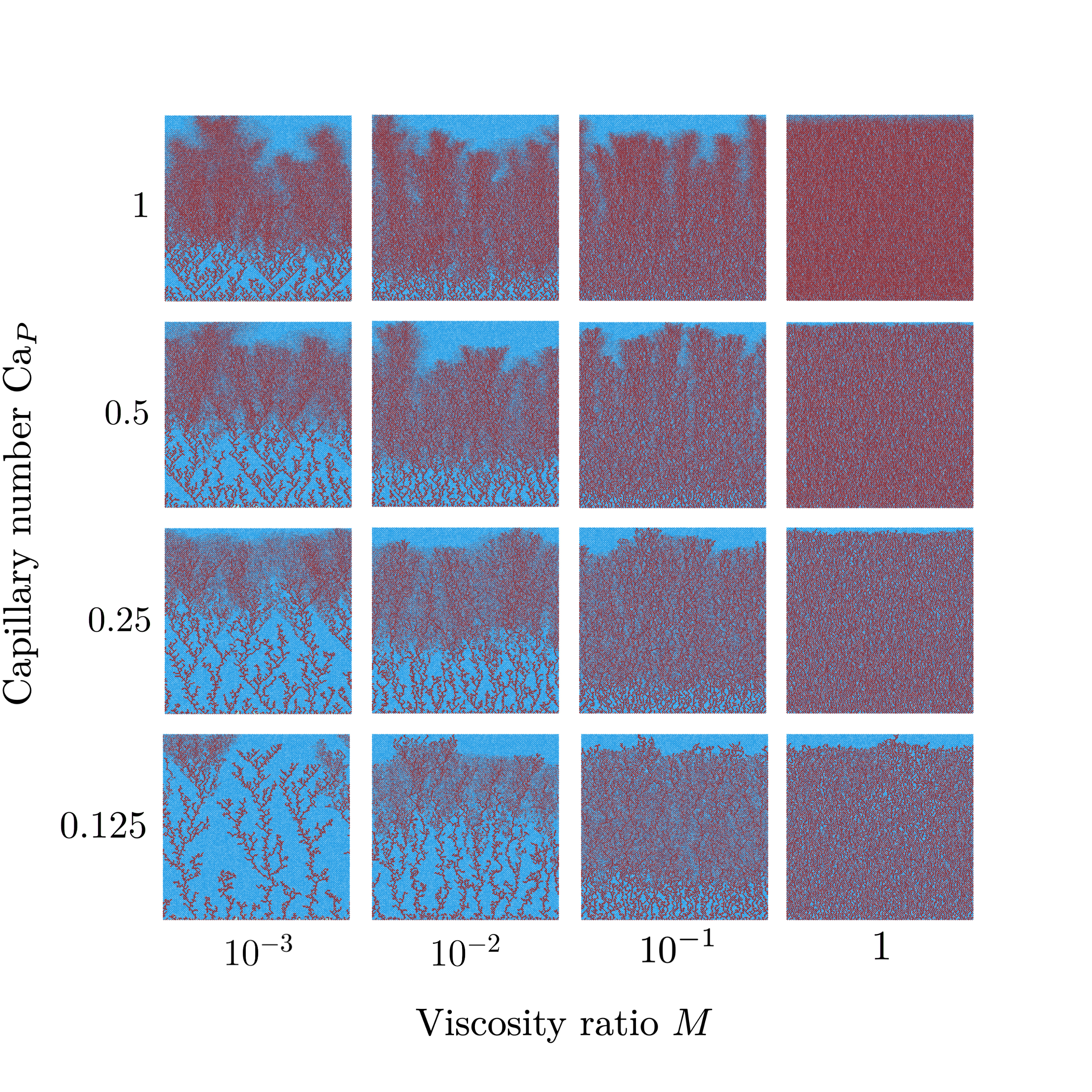}
    \caption{Diagram of the invasion patterns at breakthrough for different viscosity ratio and  capillary number. The blue and red color represent the wetting and non-wetting fluid, respectively. For these simulations $r$ is generated according to the uniform distribution \eqref{eq:Pi_uniform} with $\overline{r} = 0.25\, l$ and $a=0.15\, l$.}
    \label{fig:CaVsMdiagram}
\end{figure}

Figure \ref{fig:CaVsMdiagram} represents a diagram of the invasion pictures at breakthrough, namely when the invading fluid reaches the outlet, for different viscosity ratios and capillary numbers for which the transition is observed. It can be noted that the distance of occurrence of the transition decreases with both the capillary number and the viscosity ratio. Both trends can be qualitatively understood. As discussed above, it has been observed \cite{ransohoff88, Rossen90, dicksen02, Eriksen2022} that a certain critical pressure drop is required to generate foam. During an invasion process, we can conjecture that the foam is generated in the vicinity of the invasion front, so we should consider the pressure drops across the links close to the front and consider their evolution during the invasion. We remind that we keep the pressure drop between the inlet and the outlet fixed, and since we are injecting a fluid that has much lower viscosity than the defending fluid, the pressure gradient across the non-invaded part of the system increases as the front advances. The local pressure drops across the throats located just after the front will therefore increase as well, explaining why foam is triggered only after the front reaches a certain distance from the inlet. Moreover, the pressure gradient rises with both the parameters $\text{Ca}_P$ and $M$. Increasing either the capillary number or the viscosity ratio should then trigger the foam at a shorter distance from the inlet. One might guess that the height of the location of foam generation decreases with the inverse of both the capillary number and the viscosity ratio, namely $\Lambda \propto \text{Ca}_P^{-1}$ and $\Lambda \propto \text{M}^{-1}$. However, as we will see in the next subsection, where we present a method to characterize and analyze the transition distance $\Lambda$, this is not the case. 

\subsection{Characterization of the fingering-to-foam transition distance $\Lambda$ }
To analyse the transition from viscous fingering to foam, we first need to define a method that differentiates the two regions. For this, we define a quantity $t_{ij}=|S_{ij}-1/2|$ for every link. This is because, for a continuous displacement pattern, this quantity should be equal to $1/2$, as every link will be fully saturated by either of the two fluids and $S_{ij}$ will be either $0$ or $1$. Then, $t_{ij}$ will differ from $1/2$ for a link when it is occupied by both the fluids.  To define $\Lambda$, we then average this quantity in the direction transverse to the overall flow for a certain normalized distance from the inlet $y/l$ and find a quantity $T(y/l)$ given by
\begin{equation}
T(y/l) = \frac{1}{N_x}\sum_{x=1}^{N_x}\left|S_{ij} - \frac{1}{2}\right|.
\label{eq:T(y)}
\end{equation}
In order to decrease the noise, we average $T(y/l)$ for a given IPV over different realizations of the disorder given by different radii configurations, and we refer to this average as $\langle T(y/l) \rangle_r$. For all the simulations studied, unless specified otherwise, the average is done over 100 realizations of the radii disorder.
\begin{figure}
    \centering
    \includegraphics[scale=0.72]{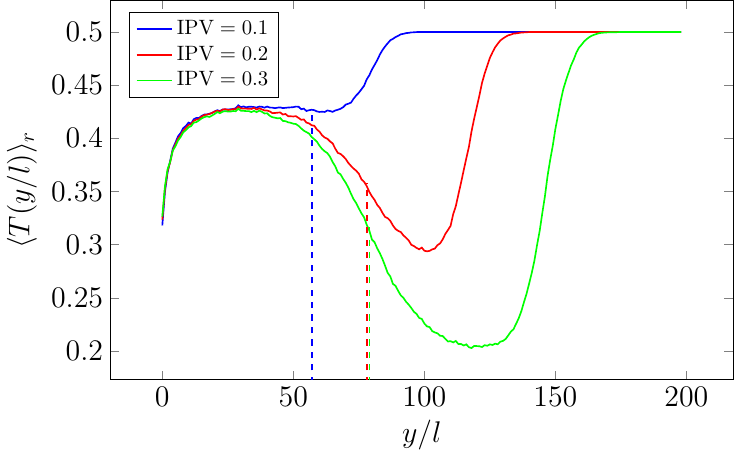}
    \caption{Plot of $\langle T(y/l) \rangle_r$, defined in equation \eqref{eq:T(y)}, for different growing IPV. For these simulations we set $\text{Ca}_P=0.25$, $M=10^{-2}$, and $r$ uniformly distributed according to equation \eqref{eq:Pi_uniform} with $\overline{r} = 0.25\,l$ and $a=0.15\,l$. Each vertical dashed line indicates the position of the minimum of the derivative of the curve of the corresponding color, which defines $\Lambda$ for that IPV.}
    \label{fig:measurement_lambda}
\end{figure}

The typical trend of $\langle T(y/l) \rangle_r$ for different IPV is shown in Figure \ref{fig:measurement_lambda}. For all IPV, and after a short distance from the inlet ($y/l \gtrsim 10$ in the figure), $\langle T(y/l) \rangle_r$ reaches a high value plateau ($\langle T(y/l) \rangle_r\simeq 0.42$ in the figure). This corresponds to the viscous fingering region, where most of the links are saturated by either one of the two fluids. Beyond this plateau, two trends are observed depending on IPV. At lower IPV, before the start of foam formation, there is no decrease in the value of $\langle T(y/l) \rangle_r$, since it increases directly to a higher plateau value 0.5. This value represents the regions fully saturated by the wetting fluid. At higher IPV, $\langle T(y/l) \rangle_r$ undergoes a significant decrease that characterizes the onset of foam, before reaching again the plateau at $0.5$.
From this curve, we define the value $\Lambda$ as the position of the minimum of the curve's derivative, as illustrated in Figure \ref{fig:measurement_lambda}.\\
\begin{figure}[ht]
    \centering
    \includegraphics[scale=0.28]{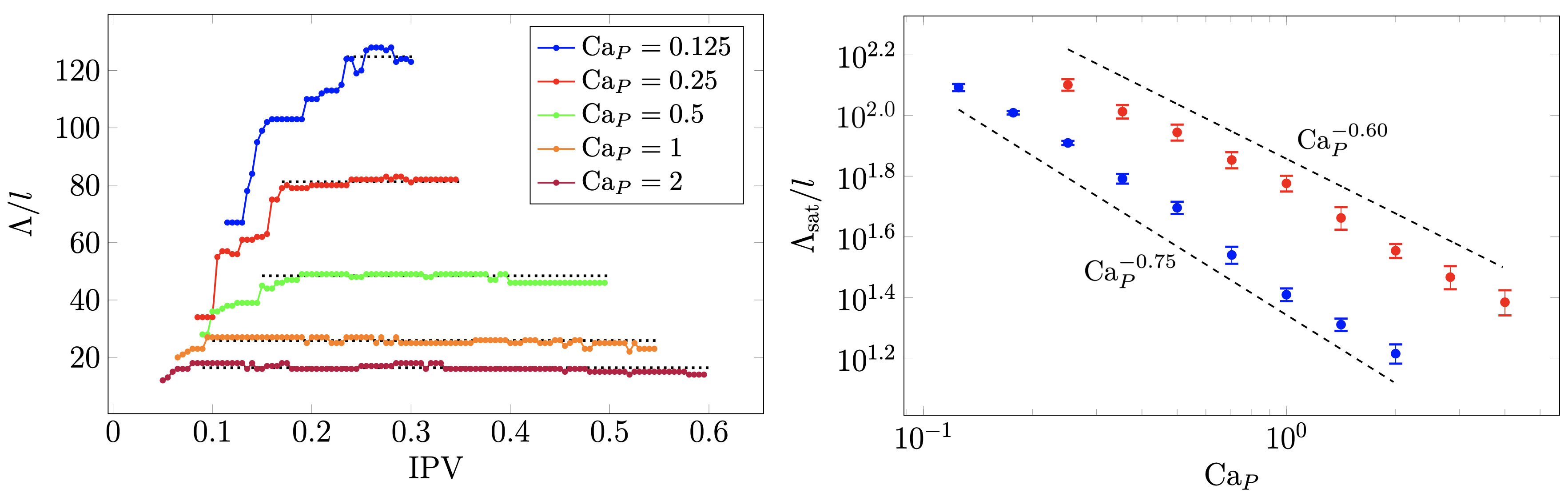}
    \caption{Left: Plot of $\Lambda/l$ as a function of IPV for different $\text{Ca}_P$, setting $M = 10^{-2}$ and uniformly distributed according to Eq. \eqref{eq:Pi_uniform} with $\overline{r} = 0.25\, l$ and $a=0.15\, l$. $\Lambda$ reaches a plateau value $\Lambda_{\text{sat}}/l$, indicated by an horizontal dotted line, which is determined by averaging the values after reaching the plateau.
    Right: $\Lambda_{\text{sat}}/l$ for different values of $\text{Ca}_{P}$. Blue dots correspond to simulations with $M=10^{-2}$ and $r$ generated according to the uniform distribution \eqref{eq:Pi_uniform} with $\overline{r} = 0.25\, l$ and $a=0.15\, l$. Red dots correspond to simulations with $M=10^{-2}$ and $r$ generated according to the Rayleigh distribution \eqref{eq:Pi_Rayleigh} with $\overline{r} = 0.1\,l$; their trend $\propto \text{Ca}_{P}^{-0.75}$ is represented by a dashed-dotted line. The error bars represent the standard deviation related to the values of $\Lambda/l$ averaged to obtain $\Lambda_{\text{sat}}/l$. The dashed lines shows the trends found with a weighted fit of the data.}
    \label{fig:lambdaVsCaP}
\end{figure}
\begin{figure}[ht]
    \centering
    \includegraphics[scale=0.28]{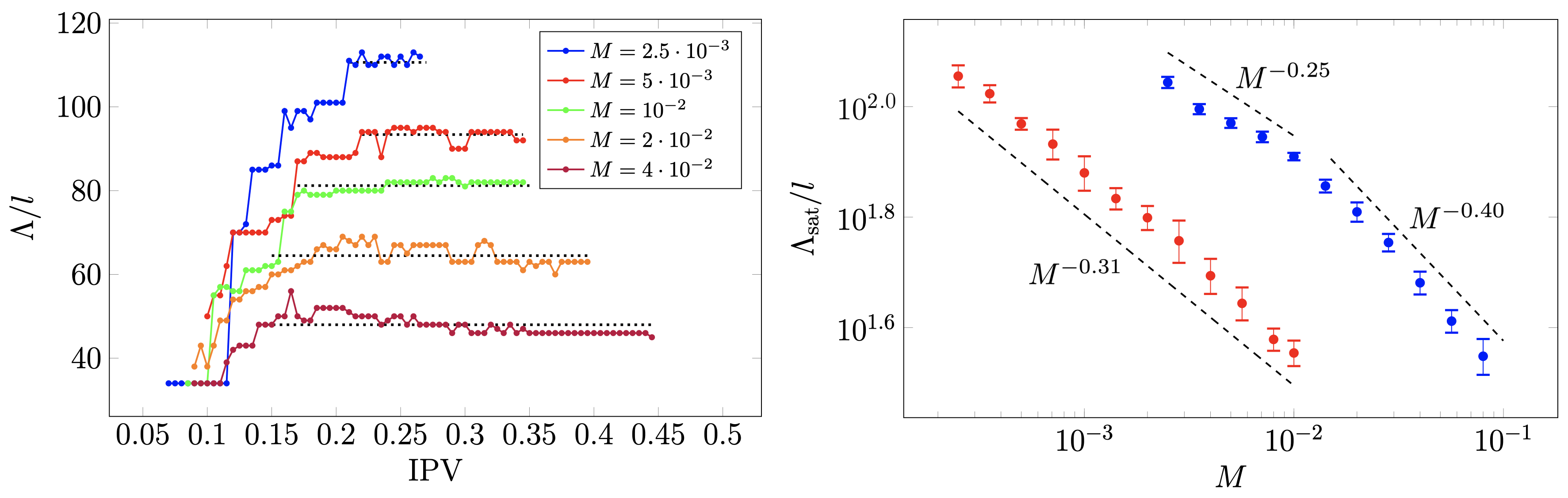}
    \caption{Left: Plot of $\Lambda/l$ as a function of IPV for different $M$, with $\text{Ca}_P = 0.25$ and $r$ uniformly distributed according to Eq. \eqref{eq:Pi_uniform} with $\overline{r} = 0.25\,l$ and $a=0.15\,l$. $\Lambda$ reaches a plateau value $\Lambda_{\text{sat}}/l$, indicated by an horizontal dotted line, which is determined by averaging the values after reaching the plateau. Right: $\Lambda_\text{sat}/l$ for different values of $M$, Blue dots correspond to simulations with $\text{Ca}_{P}=0.25$ and $r$ generated according to the uniform distribution \eqref{eq:Pi_uniform} with $\overline{r} = 0.25\,l$ and $a=0.15\,l$.  Red dots correspond to simulations with $\text{Ca}_{P}=1$ and $r$ generated according to the Rayleigh distribution \eqref{eq:Pi_Rayleigh} with $\overline{r} = 0.1\,l$. The error bars represent the standard deviation related to the values of $\Lambda/l$ averaged to obtain $\Lambda_{\text{sat}}/l$. The dashed lines shows the trends found with a weighted fit of the data. $M^{0.31}$}
    \label{fig:lambdaVsM}
\end{figure}
In Figure \ref{fig:lambdaVsCaP} (left), $\Lambda$ versus IPV is plotted for different combinations of the capillary number $\text{Ca}_P$. After an initial transient interval, $\Lambda$ reaches a plateau, where it maintains a constant average value until breakthrough. This value at the plateau depends on $\text{Ca}_P$, and in particular, for higher $\text{Ca}_P$, the $\Lambda$ plateau is lower and attained sooner. This reflects that the onset of foam formation happens at earlier stage of invasion with increasing values of $\text{Ca}_P$, as observed in Figure \ref{fig:CaVsMdiagram}. The plateau value, $\Lambda_\text{sat}$, is then calculated averaging $\Lambda$ from the end of the transient interval until the breakthrough. In Figure \ref{fig:lambdaVsCaP} (right), we plot the plateau value $\Lambda_\text{sat}$ as a function of $\text{Ca}_P$ for the two types of radii distribution, uniform and Rayleigh. We observe a non-trivial power law decay of $\Lambda_\text{sat} \propto \text{Ca}_P^{-\alpha}$. A weighted fit of the data obtained from the uniform radii distribution gives $\alpha \simeq 0.75 \pm 0.03$, while from the data from the Rayleigh radii distribution we have $\alpha \simeq 0.60 \pm 0.05$. The two trends are represented in Figure \ref{fig:lambdaVsCaP} (right) with dashed lines. It is worth noting that the simple prediction $\alpha = 1$ discussed in the previous section is invalidated.\\
The time evolution of $\Lambda$ as function of the viscosity ratio $M$ is shown in Figure \ref{fig:lambdaVsM} (left). Here also, $\Lambda$ reaches a plateau, where the value at the plateau $\Lambda_\text{sat}$ decreases with the increase of $M$ as seen in Figure \ref{fig:CaVsMdiagram}. We find that $\Lambda_\text{sat}$ follows a power-law, $\Lambda_\text{sat} \propto M^{-\beta}$. For the Rayleigh distribution, a weighted fit for all the data collected returns a power-law $\beta \simeq 0.31 \pm 0.03$. For the uniform distribution, a crossover between two different trends was observed at $M\simeq 10^{-2}$. For the data in the interval $[2.5\cdot 10^{-3},10^{-2}]$ the fit gives $\beta \simeq 0.25 \pm 0.04$, while in the interval $[\sqrt{2}\cdot 10^{-2},8\cdot 10^{-2}]$ we measured $\beta \simeq 0.40 \pm 0.06$. In Figure \ref{fig:lambdaVsM} (right) these trends are highlighted with dashed lines.\\
The independence of the power law exponents, found for both $\text{Ca}_P$ and $M$, on the radii distribution remains an open question. Recently, it was reported in literature that steady state rheology of two-phase flow depends in general on the pore size distribution. In particular, some physical quantities can vary when changing the shape of the distribution, but remaining constant if we change only the distribution width while keeping the same shape, and vice versa \cite{Roy2021}. Here, it seems that the power law is not independent from the type (so, the shape) of the radii distribution chosen, since the values of the exponents measured for the two distributions don't present a good reciprocal compatibility within the error. On the other hand, we have observed that $\Lambda$ remains constant when keeping the same type of distribution and changing only its width (see the Supplementary Material).

Now that we have quantified the location of the foam generation, in the following we investigate its influence on the total flow rate.

\subsection{Total flow rate}
\begin{figure}
    \centering
    \includegraphics[scale=0.18]{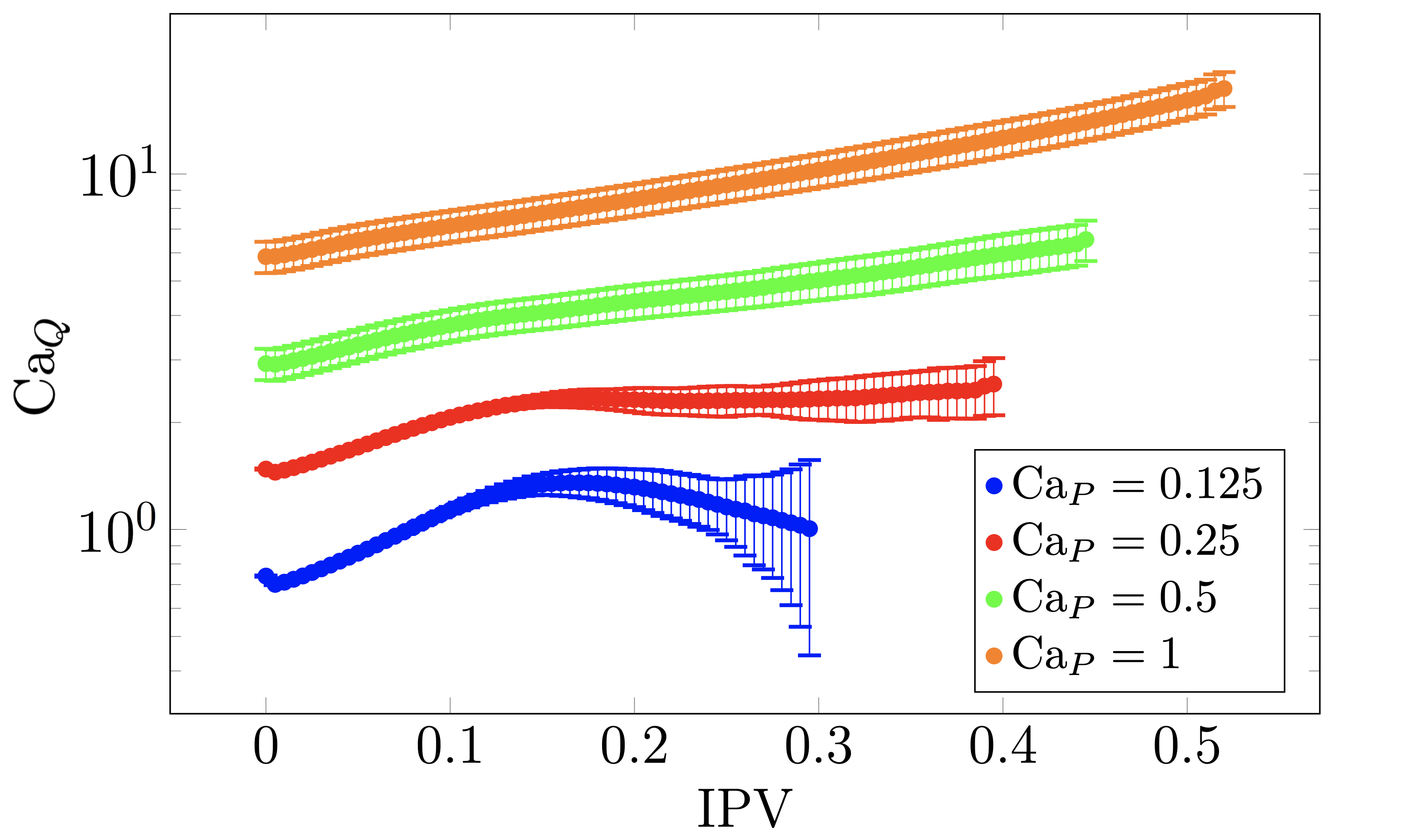}
    \caption{Plot of the dimensionless flow rate $\text{Ca}_{Q}$, defined in equation \eqref{eq:CaQ}, as a function of IPV, averaged over 1000 realizations of the radii disorder, and for different values of $\text{Ca}_P$. The error bars represent the standard deviations of the values collected from the different realizations of the disorder given by the radii configurations. For these simulations we set $M=10^{-2}$ and $r$ uniformly distributed according to equation \eqref{eq:Pi_uniform} with $\overline{r} = 0.25\,l$ and $a=0.15\,l$.}
    \label{fig:CaQtot}
\end{figure}

Figure \ref{fig:CaQtot} shows the temporal evolution of the $Q$-based capillary number $\text{Ca}_Q$, defined in equation \eqref{eq:CaQ}, for different values of $\text{Ca}_P$. Since it is directly proportional to $Q$, $\text{Ca}_Q$ can be thought as a dimensionless flow rate. Moreover, being the global pressure drop fixed throughout the evolution of the system, the total flow rate for a given IPV can be interpreted as the global permeability of the system.
For all $\text{Ca}_P$, initially the flow rate increases due to the viscous fingering instability, as we replace a more viscous fluid with a less viscous one. For later values of IPV, however, we observe an inflection point, which corresponds to the emergence of foam. This inflection is more pronounced with lower values of $\text{Ca}_P$. For the lowest capillary numbers studied, the flow rate even seems to be non-monotonic. This is an indication that the generated foam has a very low mobility, which is a typical characteristic of strong foam \cite{ransohoff88}.
Interestingly, we note that, at very low capillary number, the dispersion of the data, represented in Figure \ref{fig:CaQtot} by their standard deviation, becomes very high when the front is close to the outlet.

\begin{figure}
    \centering
    \includegraphics[scale=0.65]{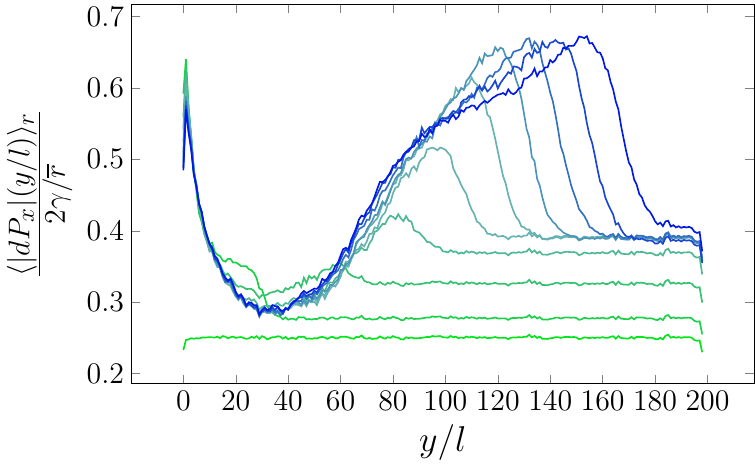}
    \includegraphics[scale=0.65]{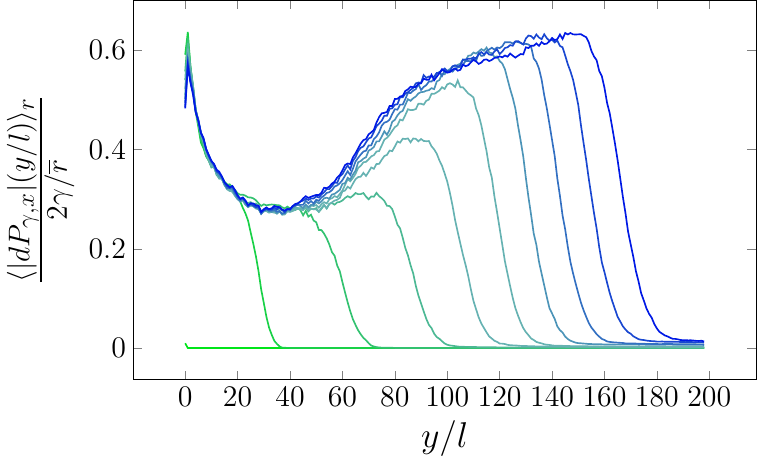}
    \caption{Left: Plot of $\langle|dP_x|\rangle_r/(2\gamma/\overline{r})$, as function of $y$-position, for different values of IPV. Different color shades represent increasing values of IPV, equally spaced from $0$ (light green) to $0.4$ (dark blue). The average is done over 100 different realization of the radii disorder. Right: Plot of the capillary contribution $\langle|dP_{\gamma,x}|\rangle_r/(2\gamma /\overline{r})$ of the pressure drop profile pictured on the right at the same IPV. For these simulations we set $\text{Ca}_P=0.25$, $M=10^{-2}$, and $r$ uniformly distributed according to equation \eqref{eq:Pi_uniform} with $\overline{r} = 0.25\,l$ and $a=0.15\,l$.}
    \label{fig:dPy}
\end{figure}

To better investigate this decrease of mobility, we looked at the evolution of the gradient of pressure along the flow direction. To do this, we average the absolute values of the local pressure drop $dP_{ij} = p_i - p_j$ along the $x$-direction:
\begin{equation}
|dP_x|(y/l) = \frac{1}{N_x} \sum_{x=1}^{N_x} |dP_{ij}|.
\end{equation}
Figure \ref{fig:dPy} (left) shows the time evolution of the normalized quantity $\langle |dP_x|\rangle_r/(2\gamma /\overline{r} )$ averaged over different realizations of the radii disorder. Close to the inlet (for $y/l\lesssim 20$), the value is quite high because of our injection condition, where many links contain a meniscus. Further, the gradient of pressure first decreases reaching a minimum, and then increases again to reach a plateau value. The lower value region corresponds to the viscous fingering region, where the pressure drop is localized in few channels of low viscosity. The higher plateau closer to the outlet corresponds instead to the region saturated with the defending viscous fluid, hence the higher value. Moreover, as discussed previously, the plateau value increases as the invasion front advances.\\
Nevertheless, the most important feature is the appearance of a bump after a certain IPV, between the minimum and the plateau, whose height and width increase as time goes by. This bump corresponds to the foam region, which decreases significantly the mobility of this region. A few observations can be made in this regard. First, the bump starts growing at approximately the same position for different IPV. The location of foam onset is thus approximately independent of time, which confirms the stability of the foam generation location. This pressure gradient measurement could thus have been used as an alternative method to quantify $\Lambda$. Second, the fact that the bump is higher than the plateau indicates that the mobility of the foam is less than that of the more viscous fluid, despite the foam being highly saturated with less viscous one. This effect is then linked to the presence of menisci in the foam. To quantify the impact of capillarity, we calculate the average of the capillary pressure drops, namely $dP_{\gamma,ij} = dP_{ij} - 8\mu_{ij} q_{ij} / (\pi r^4_{ij})$ from eq. \eqref{eq:qVSdp}, along the $x$-direction:
\begin{equation}
|dP_{\gamma,x}|(y/l)= \frac{1}{N_x} \sum_{x=1}^{N_x} |dP_{\gamma,ij}|.
\end{equation}
In Figure \ref{fig:dPy} (right) the time evolution of the normalized quantity $\langle|dP_{\gamma,x}|\rangle_r/(2\gamma /\overline{r})$ is shown. Initially, we observe a monotonic decrease between the inlet and outlet, with a sharp drop outlining the displacement front. At a certain time, a capillary pressure bump appears, which is due to foam generation and the occurrence of several menisci. We note that this bump has approximately the same magnitude and evolution as in the previous one, which confirms that mobility loss is mainly due to the presence of many menisci in this region. Another interesting feature is that the curves at different IPV seem to collapse behind the front, just before the onset of the foam, around a local minimum.

\subsection{Foam generation in the pore network}
\begin{figure}
    \centering
    \includegraphics[scale=0.34]{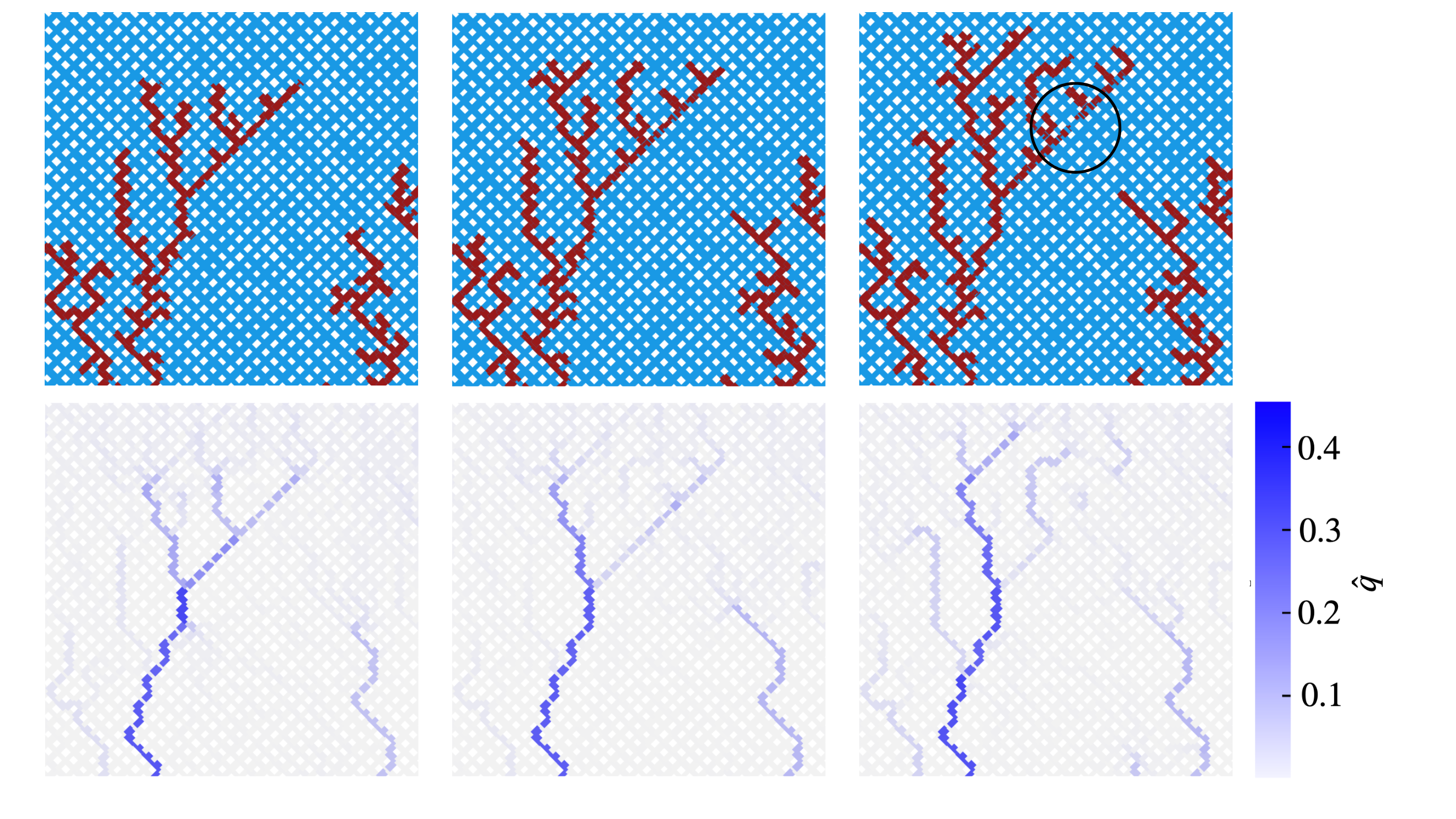}
    \caption{Foam creation in the pore network. Upper row: detail of different snapshots, at consecutive times from left to right, showing an event of foam formation. The right finger of the bigger branch fragment to create foam, highlighted with a black circle in the rightmost image. Lower row: color map of the dimensionless local flow rate $\hat{q}$ defined in equation \eqref{eq:localqhat} for the respective snapshots.}
    \label{fig:OriginOfFoam}
\end{figure}

In the previous subsection, we have quantified the occurrence of strong foam and its consequence on the total flow rate. In the introduction, we discussed a crude argument for foam generation which would occur because the gradient of pressure increases at the tip. However, from this argument one would expect that the location of $\Lambda$ decreases like $\text{Ca}_P^{-1}$ or $M^{-1}$, which is not the case. Moreover, by considering the pressure gradient for different combinations of the parameters, we were not able to identify a clear threshold value of the local pressure drop for the onset of foam. To further analyse the mechanism of foam generation, we show details of few snapshots of very early foam generation in Figure \ref{fig:OriginOfFoam} (upper row). We observe that foam is not necessarily generated at the tip of the front. Instead, blobs are produced by the fragmentation of already developed fingers located also behind the front (like for the one highlighted by the black circle in the last snapshot). In other words, branches created by the viscous fingering might be unstable and fragment at a certain location.\\
Together with this process of fragmentation, we note that there seems to be an interaction between different growing fingers. As depicted in Figure \ref{fig:OriginOfFoam} (upper row), the fragmentation of a branch is related to the approach of another (the one below on the right side). In Figure \ref{fig:OriginOfFoam} (bottom row), we present the local flow rate field corresponding to the snapshots above by plotting a dimensionless local flow rate defined as
\begin{equation}
    \hat{q} = \frac{q \mu_w}{\pi \overline{r} ^2 \gamma}
    \label{eq:localqhat}
\end{equation}
where $q$ is the local flow rate in a link. The figures show that the fragmentation of the branch occurs together with a decrease of the flow rate in the corresponding links. The flow rate in the approaching finger on the right remains instead approximately constant, so its value becomes higher than that of the breaking finger.\\
There is therefore a competition between the different branches of the invading pattern related to the emergence of foam, meaning that the fragmentation of a branch will stop its expansion, and thus favor the growth of the others. Competition between growing channel in porous media was already observed and studied, also in the context of dissolution in fractured or porous rocks \cite{hf88}. Furthermore, since the breakage of fingers cause its growth rate to drop significantly, there is a relationship between the creation and propagation of foam and the fluctuations observed in the local flow rate field. In the next subsection, we measure and characterize the time evolution of the local flow rate both in the finger region and in the foam region.

\subsection{Fluctuations of local flow rate}
\begin{figure}
    \centering
    \includegraphics[scale=0.27]{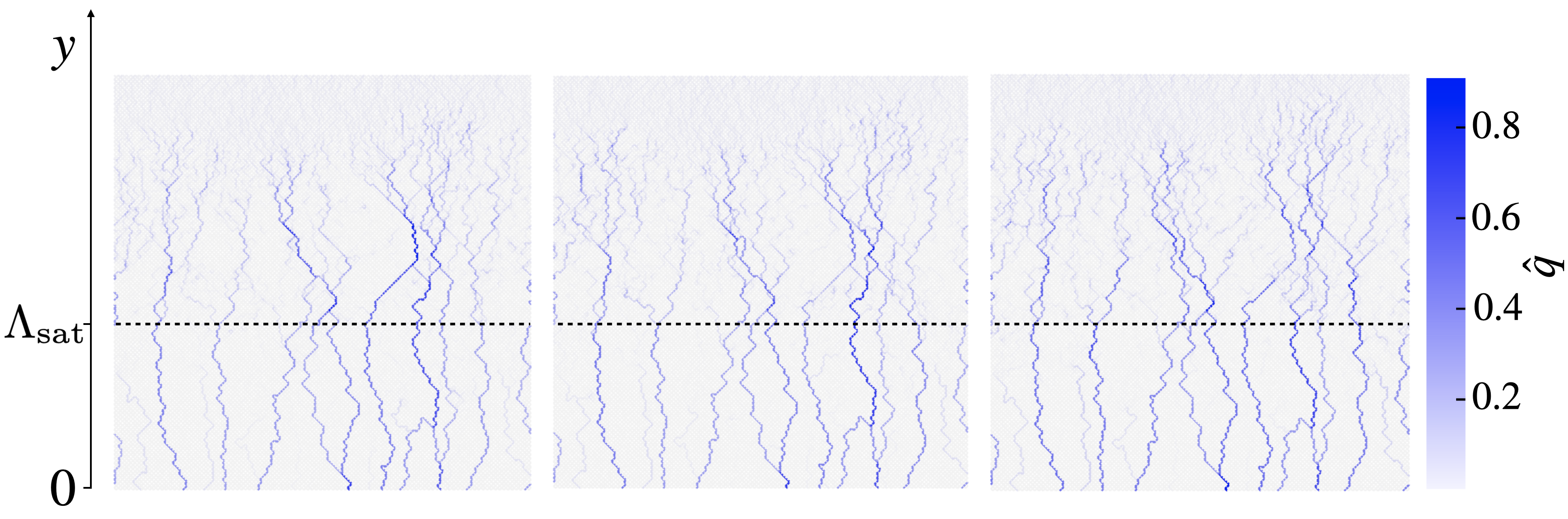}
    \caption{Color map of the dimensionless local flow rate $\hat{q}$ defined in equation \eqref{eq:localqhat}, at consecutive times from left to right, for a simulation with $\text{Ca}_P=0.25$, $M=10^{-2}$, and $r$ uniformly distributed according to equation \eqref{eq:Pi_uniform} with $\overline{r} = 0.25\, l$ and $a=0.15\, l$. The horizontal dashed lines represent the position of $\Lambda_\text{sat}$ in the pictures, corresponding to $\Lambda_\text{sat}/l \simeq 79$.}
    \label{fig:localqMap}
\end{figure}

\begin{figure}
    \centering
    \includegraphics[scale=0.34]{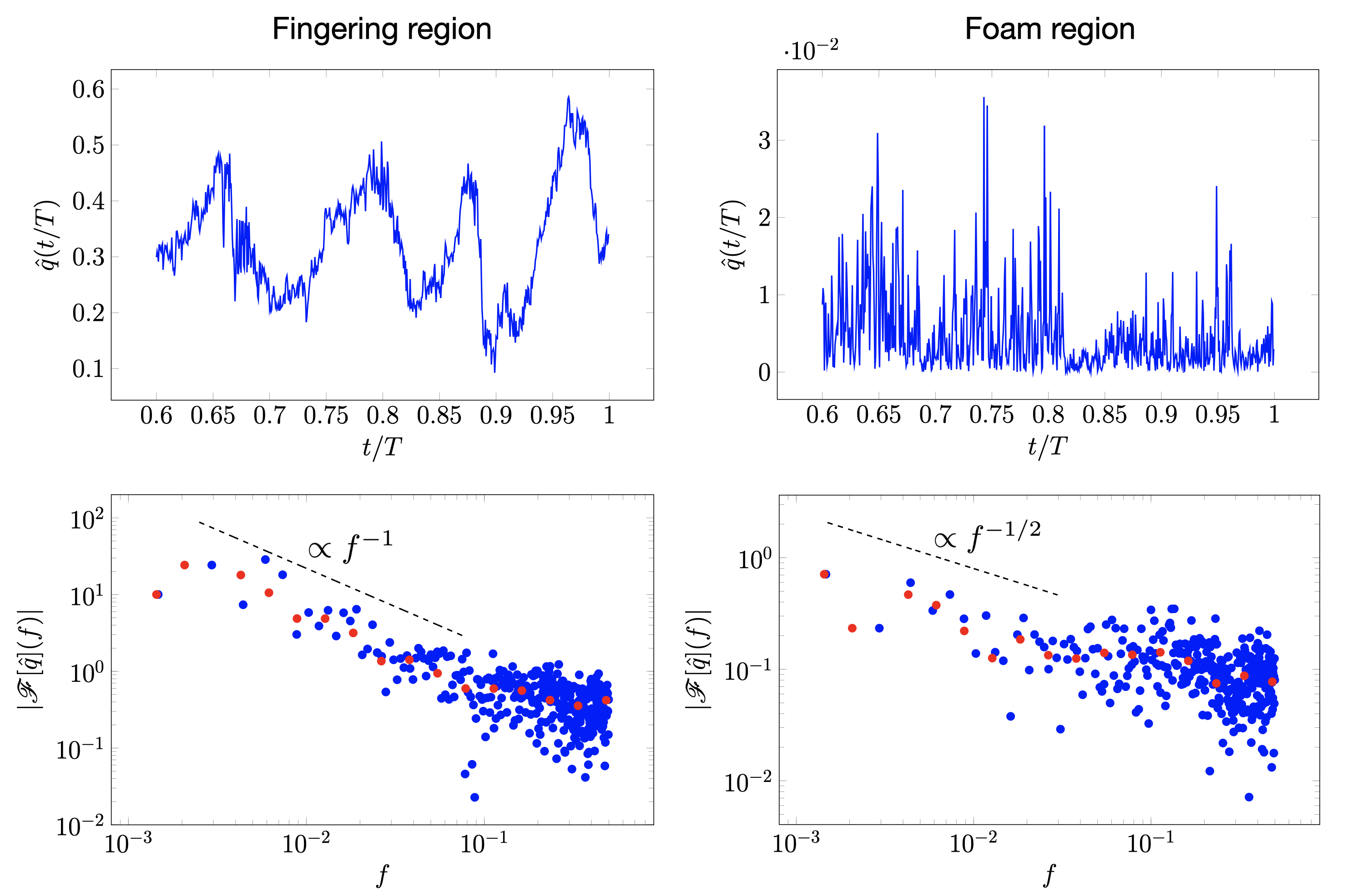}
    \caption{Upper row: Plots of the dimensionless local flow rate $\hat{q}$ defined in equation \eqref{eq:localqhat} for a link, belonging to a flowing channel, located at $y/l = 40$ (left plot) and at $y/l = 120$ (right plot), as a function of the normalized time $t/T$ in the interval $0.6 \leq t/T \leq 1$, when foam is already formed. Lower row: Plots of the absolute value of Fourier spectrum $|\mathscr{F}[\hat{q}](f)|$ of the corresponding $\hat{q}(t/T)$ shown above. Blue dots represent the values obtained from FFT. Red dots are obtained averaging these values in 20 equally logarithmic-spaced bins in the interval $[10^{-3},1]$. For this simulation, $\text{Ca}_P=0.125$, $M=10^{-2}$, and $r$ uniformly distributed according to equation \eqref{eq:Pi_uniform} with $\overline{r} = 0.25\,l$ and $a=0.15\,l$.}
    \label{fig:localqPlot}
\end{figure}

Figure \ref{fig:localqMap} illustrates the map of the dimensionless local flow rate $\hat{q}$ defined in equation \eqref{eq:localqhat} for few snapshots close to breakthrough, in a simulation setting $M=10^{-2}$, $\text{Ca}_P = 0.25$ and $r$ uniformly distributed according to equation \eqref{eq:Pi_uniform} with $\overline{r} = 0.25\, l$ and $a=0.15\, l$. As expected, in the viscous fingering region, below $\Lambda$, the flow is localized in a few, almost parallel channels, corresponding to the branches of the fingers which were not interrupted during the competition process, mentioned in the previous Section. The flow rate intensity of a single channel exhibit fluctuation in time, although the average value remains approximately stable and does not drop to zero. On the other hand, above $\Lambda$, in the foam region, the main flowing channels fragment into several smaller ones, thus distributing the flow throughout the surrounding links. In this region, new channels are continuously formed and destroyed due to the continuous foam generation led by fragmentation. It results in fluctuations of the local flow rate which are qualitatively different from the ones in the viscous fingering region.
To better investigate these fluctuations in time, we select a single link located in a flowing channel, in the fingering region and in the foam region. We plot $\hat{q}$ as a function of the normalized time $t/T$, where $T$ is the total invasion time.
Results are shown in Figure \ref{fig:localqPlot} (upper row) for a link (left) in the viscous finger region, $y<\Lambda_\text{sat}$, and a link (right) in the foam region, $y>\Lambda_\text{sat}$. 
We can see that in both cases $\hat{q}(t/T)$ resemble a stochastic process.

To characterize this stochastic process, Figure \ref{fig:localqPlot} (upper row) shows the absolute value of the temporal Fourier transform of the local flow rate, $|\mathscr{F}[\hat{q}](f)|$. We can see that, for the link at in the viscous fingering region (left), the Fourier spectrum decays approximately as the inverse of the frequency, namely $|\mathscr{F}[\hat{q}](f)|\propto f^{-1}$, for $f\lesssim 10^{-1}$. This indicates that $\hat{q}$ behaves like a random walk (Brownian) noise, at least for the lower frequencies. On the other hand, for the link in the foam region (right), the power-law decay occurs approximately as $|\mathscr{F}[\hat{q}](f)|\propto f^{-1/2}$ for $f\lesssim 10^{-1.5}$. An exponent smaller (in modulus) than $-1$ is typical of anti-correlated signals, and in particular the exponent $-1/2$ is an indicator of pink noise. As a remark, we report that we performed an analysis, not shown here, of both the local flow rate signals using wavelets, and observed results consistent with these obtained by Fourier analysis.

\section{Conclusion}
In this work, we have investigated the drainage displacement in an heterogeneous porous media. To do this, we used a dynamic pore network model which takes into account the formation of blobs. We observe that, when the flow is driven by a constant pressure difference between the inlet and the outlet, the displacement front exhibits a transition from a viscous fingering regime to a foam-like region. This transition occurs at a certain distance from the inlet, $\Lambda$, which was measured and characterized as function of the viscosity ratio $M$ and capillary number $\text{Ca}_P$. It has been shown that, after foam is generated, $\Lambda$ stabilizes to a value, $\Lambda_\text{sat}$, which decreases as a power-law for both parameters. Qualitatively, the occurrence of foam can be explained from the fact that the pressure gradient is not homogeneous through the medium, but is stronger in the defending fluid and is increasing as the front advances. This pressure gradient, in competition with capillary force, might trigger fragmentation and thus foam formation. From this qualitative argument, we cannot however predict the observed exponents for the power-laws decay. Further analysis would be thus necessary to understand this transition.

Moreover, we have shown that foam formation could be related to an instability mechanism, with fragmentation of the viscous fingers occurring below the front. We might remark that foam is also responsible for an increase in the local pressure gradient, which in turn should further promote creation of more foam. Furthermore, we observe that the flow velocity is not homogeneous but localized in few channels, both in the viscous fingers and in the foam region, although in the latter the channels tends to divide into several smaller ones. This leads to very different behaviour of the local flow rate in the fingers region, where the fluctuations are mostly stable in time, from the foam region, where the value drops intermittently to zero. The Fourier analysis suggest that the first resemble a Brownian motion, while the second an anti-correlated (pink) signal. It is clear that the two behaviours are causally connected, although we still don't understand how the fluctuations in the foam, driven by the fragmentation process, influence the evolution observed in the fingering region, and, eventually, vice versa. However, we can add that the strong fluctuations observed in the foam region can be responsible of the irregular movement of the front observed when foam is formed and is propagating.

Future research efforts can be dedicated to the experimental observation of the transition and the validation of the predictions elaborated in this work, conceivably building a setup analogous to the one adopted in \cite{Eriksen2022}, in a rectangular geometry. Furthermore, the Pore Network Model presented here can be extended to a three-dimensional domain for comparison with the two-dimensional analysis reported in this study. It is also worth considering the study of two-phase flows where one of the two fluids is non-Newtonian, and evaluate the impact of the non-linear rheology, like the presence of yield stress, in foam formation and propagation.

\section*{Declarations}
\textbf{Author Contributions}\ FL wrote the code specific for this project based on algorithms written by LT and SS. FL performed the simulations and analyzed the data. LT helped with the data analysis and with the study of the total flow rate and of the local flow rate fluctuations. SS helped with the study of the characterization of the transition distance. AH and LT suggested the problem and the interpretation of the foam generation. AR supervised the whole work. All the authors contributed in writing the manuscript to its final form.\\

\textbf{Funding}\ This work was partly supported by the Research Council of Norway through its Center of Excellence funding scheme, project number 262644. Further support, also from the Research Council of Norway, was provided through its INTPART program, project number 309139. This work was also supported by ”Investissements d’Avenir du LabEx” PALM (ANR-10-LABX-0039-PALM).\\

\textbf{Acknowledgment}\ We thank A. Andersen Hennig and V. M. Schimmenti for discussions and suggestions.

\let\oldaddcontentsline\addcontentsline
\renewcommand{\addcontentsline}[3]{}
\bibliographystyle{apsrev4-1}
\bibliography{biblio}
\let\addcontentsline\oldaddcontentsline

\end{document}


\onecolumngrid

\graphicspath{{SuppFigures/}}

\newpage 
	\setcounter{equation}{0}
	\begin{center}
		{\Large Supplementary Material \\
			Dynamic pore-network modeling of transition from viscous fingers to compact displacement during drainage
		}
	\end{center}

\section{Dependence of the invasion pattern from the parameters}
\begin{figure}[b]
    \centering
    \includegraphics[scale=0.14]{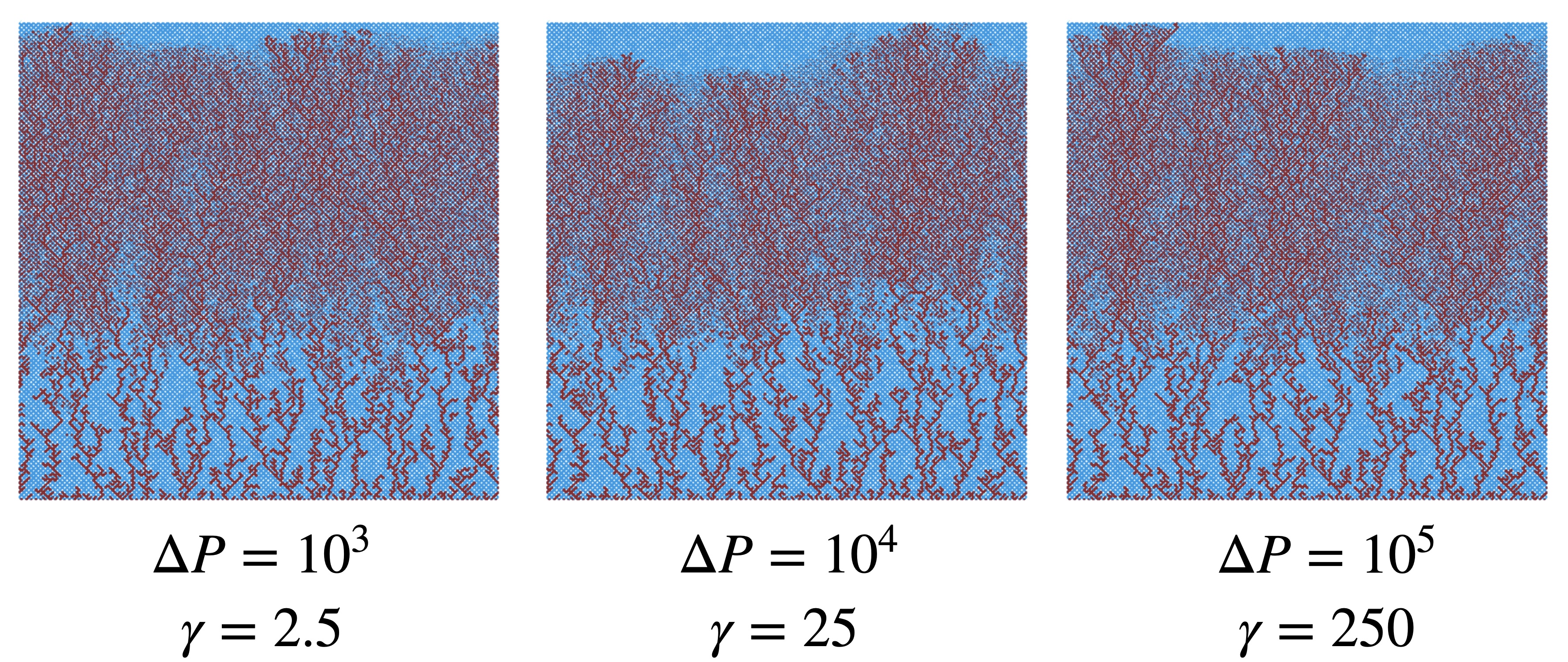}
    \caption{Snapshots of the invasion pattern at breakthrough setting different combinations of the surface tension $\gamma$ and the global pressure drop $\Delta P$ but keeping $\text{Ca}_P=0.25$ For these simulations we set $N_x = N_y = 200$, $M=10^{-2}$ and $r$ uniformly distributed according to eq. \eqref{eq:Pi_uniform_supp} with $\overline{r} = 0.25l$ and $a=0.15l$.}
    \label{fig:varying_gamma}
\end{figure}

\begin{figure}
    \centering
    \includegraphics[scale=0.14]{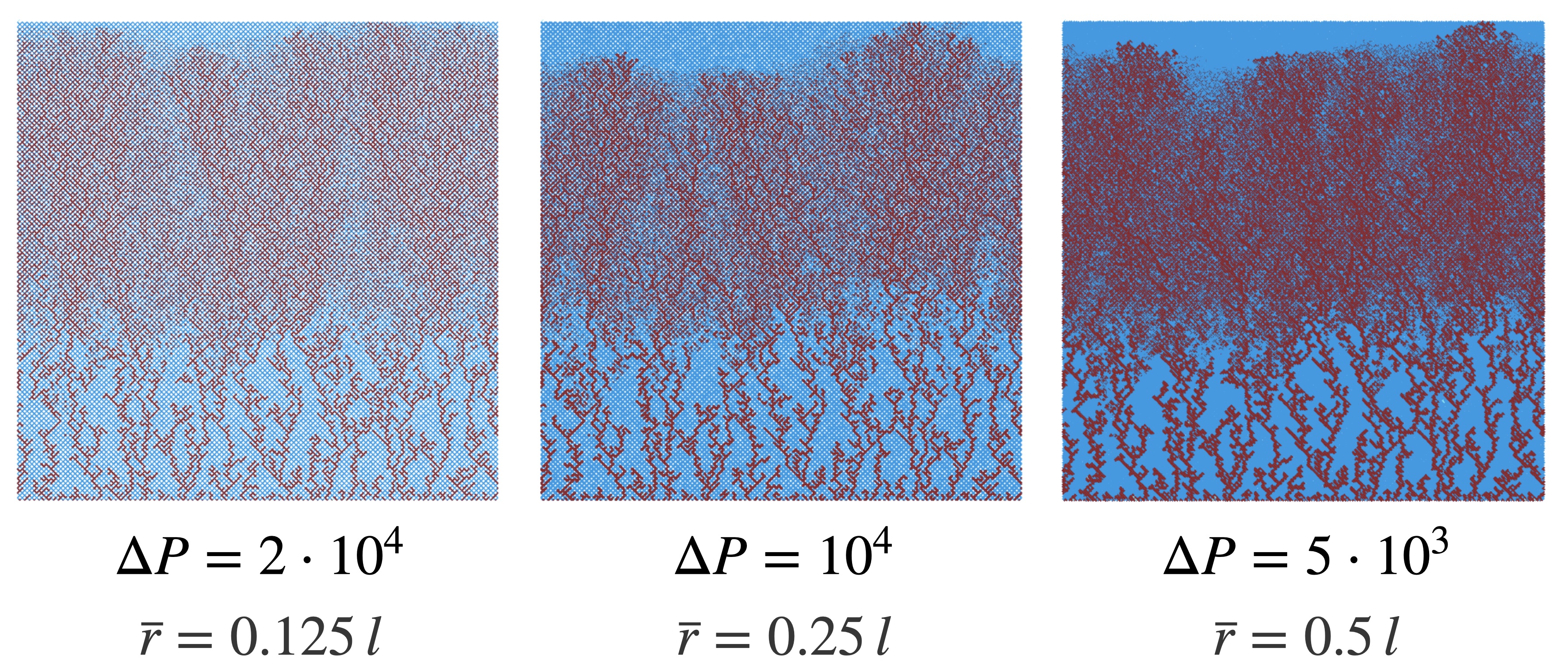}
    \caption{Snapshots of the invasion pattern at breakthrough setting different combinations of the average radius $\overline{r}$ and the global pressure drop $\Delta P$ but keeping $\text{Ca}_P=0.25$. For these simulations we set $N_x = N_y = 200$, $M=10^{-2}$ and $r$ uniformly distributed according to eq. \eqref{eq:Pi_uniform_supp} with $a=0.15l$.}
    \label{fig:varying_r}
\end{figure}

\begin{figure}
    \centering
    \includegraphics[scale=0.14]{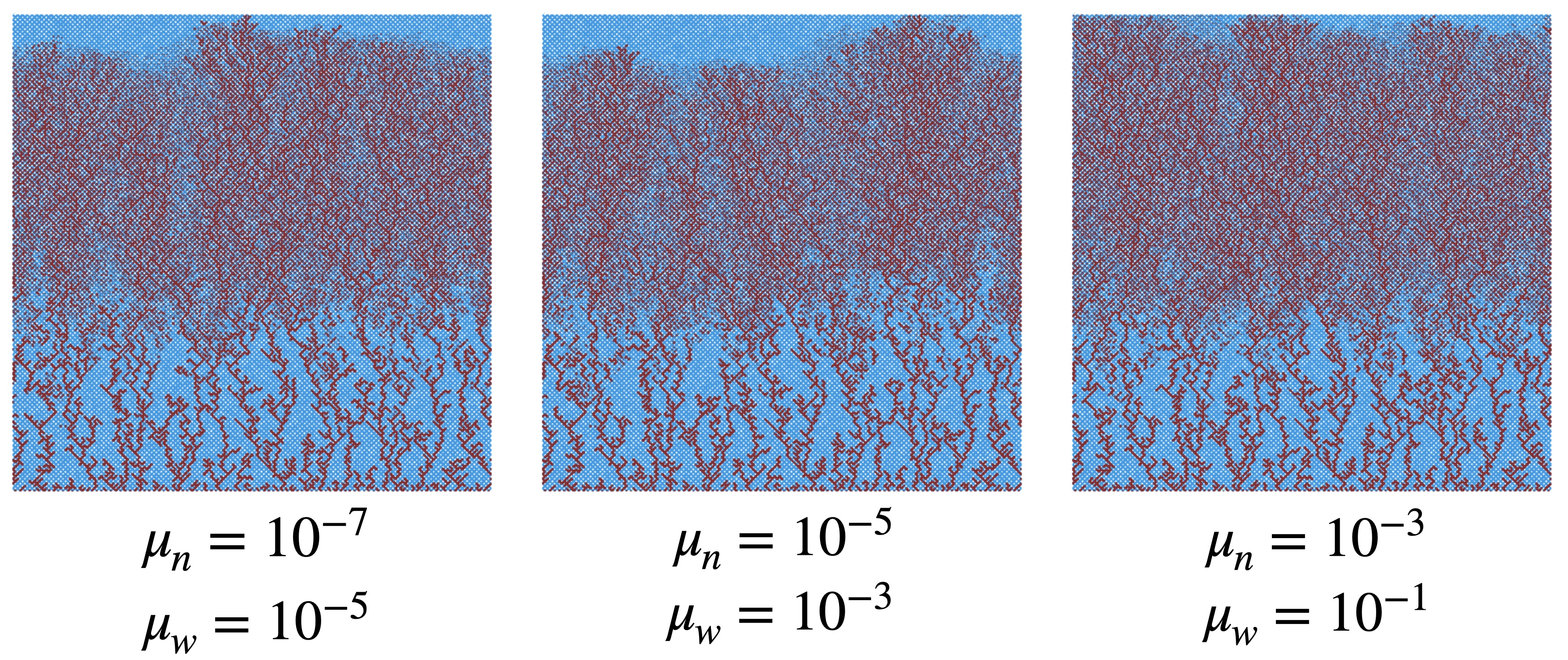}
    \caption{Snapshots of the invasion pattern at breakthrough setting different combinations of the viscosities $\mu_n$ and $\mu_w$ but keeping $M=10^{-2}$. For these simulations we set $N_x = N_y = 200$, $\text{Ca}_P=0.25$ and $r$ uniformly distributed according to eq. \eqref{eq:Pi_uniform_supp} with $\overline{r} = 0.25l$ and $a=0.15l$.}
    \label{fig:varying_mu}
\end{figure}

In our work, we assumed that the viscosity ratio $M = \mu_N/\mu_W$ and the capillary number $\text{Ca}_P = (\Delta P/L_x) / (2\gamma/\overline{r})$ fully characterize the displacement pattern occurring in our system. In this Section of the Supplementary Material, we want to verify this assumption, showing that different simulations prepared setting the same values for $M$ and $\text{Ca}_P$ will return the same pattern. For each of the invasion patterns shown in Figure \ref{fig:varying_gamma}, different values for both the surface tension $\gamma$ and the global pressure drop $\Delta P$ were chosen, but such that $\text{Ca}_P$ is the same \footnote{The unit of measure of the physical quantities, not reported here, can be chosen arbitrarily, as long as the choice remain consistent with the fact that the global parameters introduced must be dimensionless.}. The patterns are very similar to each other, varying only by small details at the scale of few pores. In particular, the foam originates and propagates from the same transition height $\Lambda$. Analogously, Figure \ref{fig:varying_r} show three different patterns for which both the average radius $\langle r \rangle$ and $\Delta P/L_x$ are varied. Also in this case, identical values of $\text{Ca}_P$ leads to similar patterns. An analogue test for the viscosity ratio $M$ was done. In Figure \ref{fig:varying_mu} it is shown that a different choice for the wetting and non-wetting viscosities, $\mu_w$ and $\mu_n$, will not modify the macroscopic output, as long as their ratio is kept equal.

\begin{figure}
    \centering
    \includegraphics[scale=0.14]{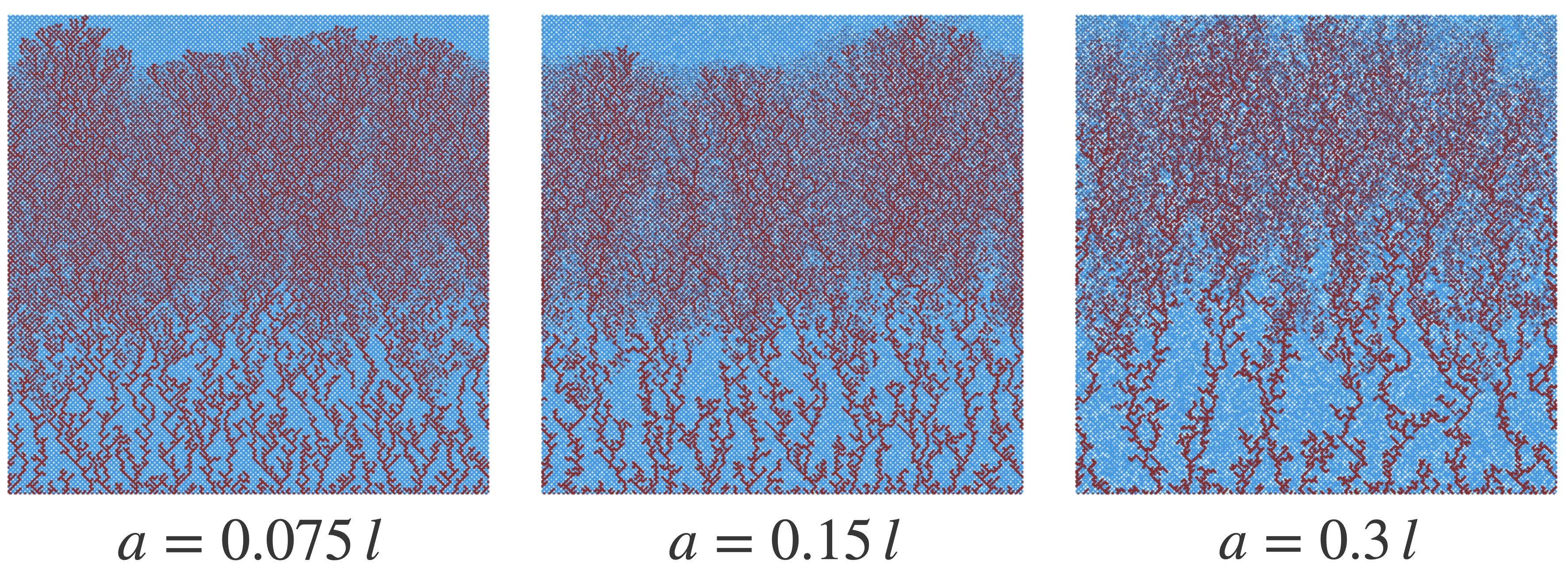}
    \caption{Snapshots of the invasion pattern at breakthrough with $r$ uniformly distributed according to eq. \eqref{eq:Pi_uniform_supp} with $\langle r \rangle = 0.25l$ and different values for $a$. For these simulations we set $N_x = N_y = 200$, $\text{Ca}_P=0.25$ and $M=10^{-2}$. }
    \label{fig:varying_a}
\end{figure}

We also investigate the dependence of the invasion pattern from the width of the radii distribution. For different simulations, we generate the tubes radii from a uniform distribution
\begin{equation}
    \Pi(r) =\begin{dcases} 1/a & \text{if}\; r\in \left[ \overline{r} - a/2, \overline{r} + a/2 \right]\\
0 & \text{otherwise}
\end{dcases}\!,
\label{eq:Pi_uniform_supp}
\end{equation}
having the same $\overline{r}$ but different width $a$. The outputs are presented in Figure \ref{fig:varying_a}, where we observe $\Lambda$ to remain the same, while the number of finger and the compactness of the foam seem both to decrease as $a$ rises. A further analysis might characterize and quantify this relationship.

\begin{figure}
    \centering
    \includegraphics[scale=0.155]{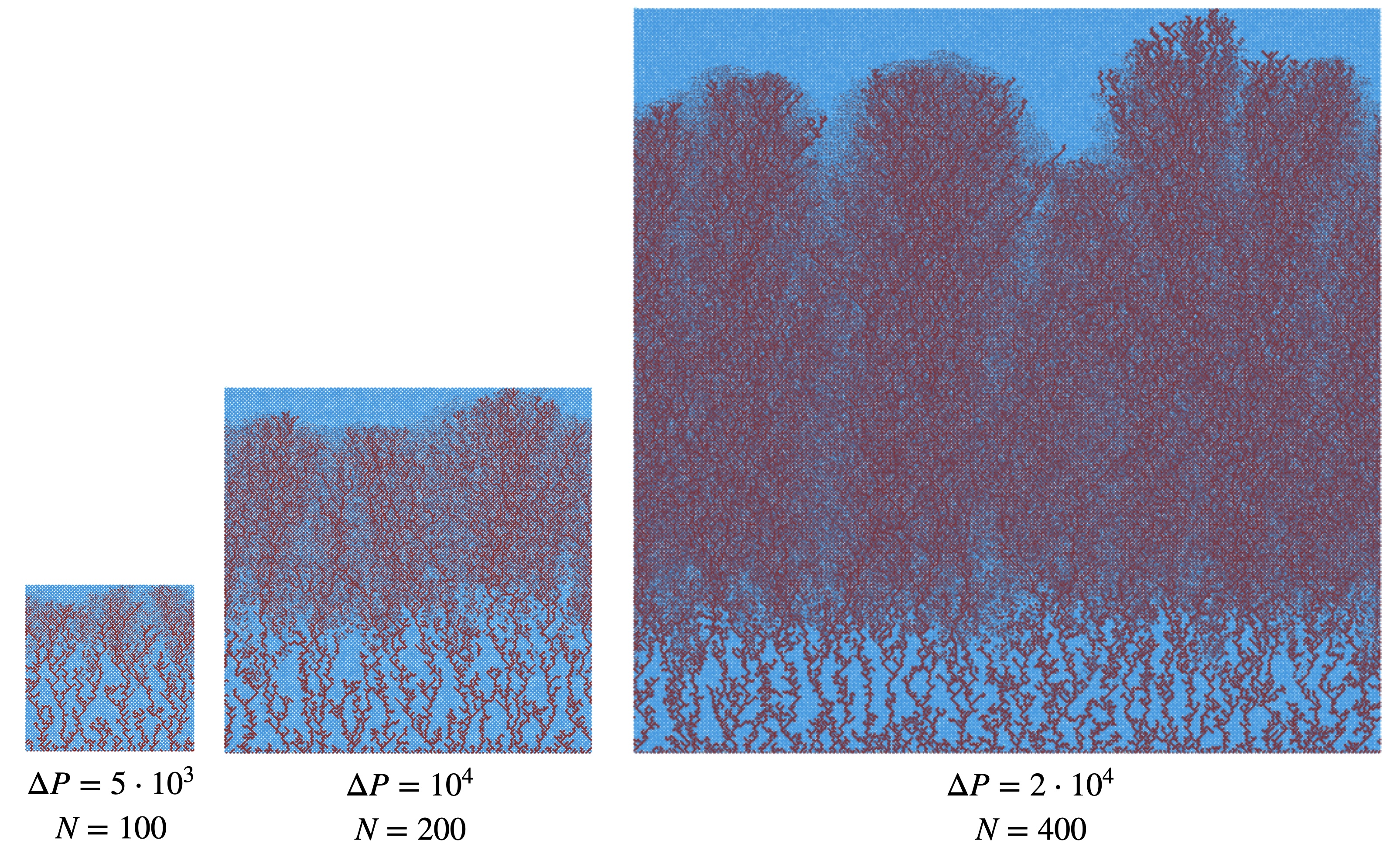}
    \caption{Snapshots of the invasion pattern at breakthrough setting different system lengths $N = N_x = N_y$. For these simulations we set $\text{Ca}_P=0.25$, $M=10^{-2}$, and $r$ uniformly distributed according to eq. \eqref{eq:Pi_uniform_supp} with $\overline{r} = 0.25l$ and $a=0.15l$.). }
    \label{fig:varying_L}
\end{figure}

Finally, to look at the influence of the network size, in Figure \ref{fig:varying_L} we show some patterns obtained varying only the number of tubes per side, $N_x = N_y \equiv N$ in a square network. We observe the finger-foam transition to occur at approximately the same distance $\Lambda$ from the inlet, so the foam just propagates for longer distances when $N$ is bigger.


